\documentclass[aps,prd,preprint,a4paper,showpacs,nofootinbib,superscriptaddress]{revtex4-2}
\usepackage{bm}
\usepackage{indentfirst}
\usepackage{amsmath}
\usepackage{graphicx}
\usepackage{amssymb}
\usepackage{subfigure}
\usepackage{amssymb}
\usepackage{hyperref}
\usepackage{epstopdf}
\usepackage{cancel}
\usepackage{booktabs}
\usepackage[section]{placeins}

\usepackage[utf8]{inputenc}
\hypersetup{
    colorlinks=true,
    linkcolor=red,
    citecolor=blue,
}
\usepackage{color}
\usepackage[T1]{fontenc}
\usepackage{txfonts}
\usepackage{orcidlink}
\usepackage[title]{appendix}	

\begin{document}

\title{Novel black holes with scalar hair in the Einstein-Maxwell-Scalar Theory with positive coupling}

\author{Hong Guo}
\email{hong_guo@usp.br}
\affiliation{Escola de Engenharia de Lorena, Universidade de São Paulo, 12602-810, Lorena, SP, Brazil}

\author{Wei-Liang Qian}
\email{wlqian@usp.br}
\affiliation{Escola de Engenharia de Lorena, Universidade de São Paulo, 12602-810, Lorena, SP, Brazil}
\affiliation{Faculdade de Engenharia de Guaratinguet\'a, Universidade Estadual Paulista, 12516-410, Guaratinguet\'a, SP, Brazil}
\affiliation{Center for Gravitation and Cosmology, College of Physical Science and Technology, Yangzhou University, 225009, Yangzhou, China}

\author{Bean Wang}
%\email{bin.wang@vanguard.edu}
\affiliation{Department of Physical Sciences and Applied Mathematics, Vanguard University, Costa Mesa, CA 92626, USA}

\begin{abstract}
In this work, we find a new branch of hairy black hole solutions in the Einstein–Maxwell–Scalar (EMS) theory in four-dimensional asymptotically flat spacetimes.
Different from spontaneous scalarization induced by tachyonic instabilities in Reissner–Nordstr\"{o}m (RN) black holes with a negative coupling parameter, these scalar-hairy black hole solutions arise when the coupling parameter is positive, where nonlinear coupling plays the dominant role, meaning that the coupling is positively correlated with the degree of deviation from the trivial state.
Our numerical analysis reveals that the scalar field grows monotonically with the radial coordinate and asymptotically approaches a finite constant, exhibiting behavior that is qualitatively similar to that of the Maxwell potential.
In these solutions, an increase in the charge $q$ causes the scalar-hairy solutions to deviate further from the RN state, while excessive charging drives the system back towards hairless solutions.
Strengthening the coupling parameter compresses the existence domain of the scalar-hairy state, which lies entirely within the parameter region of RN black holes.
Moreover, by evaluating the quasinormal modes, we show that the obtained scalar-hairy solutions are stable against linearized scalar perturbations.
\end{abstract}

\maketitle

\newpage

%%%%%%%%%%%%%%%%%%%%%%%%%%%Section%%%%%%%%%%%%%%%%%%%%%%%%%%%
\section{Introduction}\label{sec=intro}

Research on black holes plays a pivotal role in advancing gravitational theories, cosmology, and astronomy, primarily serving as a fertile testing ground for general relativity (GR) and alternative theories of gravity.
The remarkable success of GR in astrophysical and cosmological testing has profoundly shaped our understanding of gravitational phenomena.
To date, it has passed several precise observational tests~\cite{Will:2014kxa,Yagi:2016jml} across multiple scales: laboratory experiments~\cite{Adelberger:2003zx}, solar system verifications~\cite{Sakstein:2014isa,Ni:2016dwy}, astrophysical systems tests~\cite{Berti:2015itd,Ishak:2018his}, gravitational wave detections from compact mergers~\cite{LIGOScientific:2017vwq,LIGOScientific:2018mvr,LIGOScientific:2019fpa}, and future probes using extreme mass-ratio inspirals (EMRIs) with space-borne gravitational wave detectors~\cite{Berry:2019wgg}.

The advancement of astronomical detection techniques has simultaneously driven significant interest in modified gravity theories~\cite{Clifton:2011jh}.
A cornerstone of GR is the no-hair theorem, which states that stationary black holes in Einstein-Maxwell theory must belong to the Kerr-Newman family, characterized solely by mass, charge, and angular momentum~\cite{Kerr:1963ud,Bekenstein:1972ky,Teitelboim:1972qx,Chrusciel:2012jk}.
However, studies have revealed that black holes in modified gravity theories can possess macroscopic external degrees of freedom.
These include coupling with matter fields~\cite{Winstanley:1998sn}, alterations in the asymptotic structure of spacetime~\cite{Ponglertsakul:2016wae}, or entirely alternative theoretical frameworks beyond Einstein–Maxwell theory~\cite{Brihaye:2015qtu}.
The deviations of black hole solutions in modified gravity from those in pure GR serve as important tools for distinguishing extra fields and various GR extensions, as they exhibit richer physics than their bald counterparts~\cite{Nunez:1996xv,Ghosh:2025igz}. 
These deviations are predominantly manifested in the form of black hole hairs~\cite{Coleman:1991ku,Herdeiro:2015waa,Volkov:2016ehx}.

In four-dimensional flat spacetime, the first hairy black hole solution was known as the BBMB black hole~\cite{Bekenstein:1974sf,Bekenstein:1975ts}, which incorporated GR and a conformally coupled scalar field.
However, this solution was soon found to exhibit instability under scalar perturbation~\cite{Bronnikov:1978mx}.
Subsequently, static hairy black holes in SU(2) Einstein-Yang-Mills theory became more widely recognized~\cite{Volkov:1989fi,Volkov:1990sva,Bizon:1990sr}. 
This discovery was subsequently extended in various ways, leading to numerous black hole solutions with Skyrmion hair~\cite{Luckock:1986tr,Droz:1991cx,Bizon:1992gb}, Higgs hair~\cite{Greene:1992fw}, stringy hair~\cite{Kanti:1995vq}, axionic hairs~\cite{Campbell:1990ai}, non-spherically symmetric hair~\cite{Kleihaus:1997ic}, as well as black hole models carrying magnetic monopoles~\cite{Breitenlohner:1991aa,Breitenlohner:1994di,Greene:1992fw}. 
For further details, refer to the review ~\cite{Volkov:1998cc}, with recent developments~\cite{Sotiriou:2013qea,Herdeiro:2014goa,Brito:2013xaa,Herdeiro:2016tmi,Hong:2020miv,Herdeiro:2020xmb}.

Notably, black holes with scalar hair play a crucial role in circumventing the no-hair theorem.
As the simplest matter field, scalar fields have become pivotal in various aspects of black hole physics and cosmology. 
They feature prominently in dark matter models~\cite{Marsh:2015xka,Cardoso:2019rvt}, inflation of cosmology~\cite{Bezrukov:2007ep,Barbon:2009ya,Lyth:1998xn}, gravitational collapse of scalar fields and black hole formation~\cite{Choptuik:1992jv,Goncalves:1997qp}.
Moreover, through conformal transformations, scalar fields provide the connections between GR and modified gravity theories~\cite{Wagoner:1970vr,DeFelice:2010aj,Sotiriou:2008rp,Nojiri:2010wj,Nojiri:2017ncd}, which constitutes a primary motivation for investigating black hole solutions with additional scalar hair.
Further details can be found in~\cite{Herdeiro:2015waa,Volkov:2016ehx}.
Recent studies have revealed several intriguing mechanisms for black holes with scalar hair.
In Kerr or charged black hole spacetimes, time-dependent bosonic fields can trigger superradiant instabilities~\cite{Press:1972zz,Sanchis-Gual:2015lje,Brito:2015oca}, leading to the synchronization of the black hole~\cite{Herdeiro:2014goa,East:2017ovw,Herdeiro:2017phl} and ultimately to the formation of hairy black hole solutions.
Simultaneously, studies demonstrated that a nonminimal coupling between the scalar field and the Gauss–Bonnet curvature can induce tachyonic instabilities in vacuum black holes, giving rise to spontaneous scalarization of black holes~\cite{Doneva:2017bvd,Silva:2017uqg,Antoniou:2017acq,Herdeiro:2018wub}.
A more straightforward approach revealed that Reissner–Nordstr\"{o}m (RN) black holes become unstable when the scalar field is nonminimally coupled to the electromagnetic field, leading to the black hole spontaneous scalarization in the Einstein–Maxwell–scalar (EMS) theory~\cite{Herdeiro:2018wub,Fernandes:2019rez}.
The stability of these scalarized black holes and their quasinormal modes (QNM) have been extensively analyzed in various studies~\cite{Myung:2019oua,Myung:2018jvi,Zou:2020zxq,LuisBlazquez-Salcedo:2020rqp}.

In fact, in asymptotically anti-de Sitter (AdS) spacetime, holographic superconductors constitute a distinct class of hairy black holes with scalar hair.
As elaborated in~\cite{Hartnoll:2008kx,Hartnoll:2008vx}, when the system temperature drops below a critical value, spontaneous symmetry breaking occurs.
This manifests holographically as the scalar field acting as a vacuum expectation value begins to emerge in the boundary field theory, thereby simulating superconducting phase transitions. 
In the bulk, the original AdS black hole undergoes tachyonic instability, leading to the formation of a new black hole solution with scalar hair~\cite{Gubser:2008px}. 
Another well-known model in holographic theories is the holographic QCD model~\cite{DeWolfe:2010he,Kim:2012ey,Dudal:2017max,Rougemont:2023gfz}.
Recently, an improved Einstein–Maxwell–dilaton (EMD) model~\cite{Critelli:2017oub,Grefa:2021qvt,Cai:2022omk} has been proposed in which, via the AdS/CFT correspondence, the scalar field in the boundary theory acts as a scaling factor for the energy scale~\cite{Cai:2022omk}, successfully simulating the expected QCD phase transition and the existence of a critical endpoint. 
Here, the gauge field in the boundary theory corresponds to the baryon density and baryon chemical potential of the QCD model, while the black hole’s Hawking temperature and Bekenstein entropy holographically map to the temperature and entropy density, respectively.
Clearly, the bulk theory necessarily contains a nontrivial scalar field to describe holographic QCD dynamics.

While previous studies have largely focused on modeling the QCD phase transition, understanding the gravitational sector of EMD models remains essential for understanding the nature of the hairy solutions and the associated black hole state transitions that arise in the bulk.
Notably, the relevant degrees of freedom of EMS and holographic EMD theories closely resemble each other, with action
\begin{equation}
	S=\int d x^4\left(R-\frac{1}{2} \nabla^\mu \psi \nabla_\mu \psi-\frac{1}{4} f(\psi) F_{\mu \nu}^2-V(\psi)\right).
\end{equation}
The EMD model employs more intricate functions $f(\psi)$ and $V(\psi)$, whereas the original EMS theory was formulated with a massless real scalar field as $V(\psi)=0$.
In~\cite{Guo:2024ymo}, using a massive real scalar field EMS model in asymptotically AdS spacetime, we obtained phase diagrams qualitatively similar to those in~\cite{Critelli:2017oub,Grefa:2021qvt}, identifying a first-order phase transition below a critical point.
Furthermore, we revealed a third-order transition above the critical point by demonstrating the coexistence of two distinct scalarized black hole states in the bulk, enabling precise determination of phase boundaries in the dual field theory.
This suggests a close connection between hairy black holes in EMD and EMS theories.
However, under the weak-field approximation, the coupling function $f(\psi)$ in the EMD theory corresponds to a positive coupling in the EMS theory, implying these hairy solutions don't originate from spontaneous scalarization.
As discussed in~\cite{Herdeiro:2018wub,Fernandes:2019rez}, spontaneous scalarization in EMS theory requires $\partial^2_{\psi}f(\psi)>0$ and $\psi\partial_{\psi}f(\psi)<0$ for tachyonic instability.
For a coupling of the form $f(\psi)=e^{-\lambda\psi^2}$, these conditions demand a negative coupling parameter $\lambda<0$.
These findings collectively indicate more than one type of hairy black hole solution exists in EMS theory.

Opposite coupling coefficients in spontaneous scalarization have also garnered considerable research interest.
Similar to neutron star scalarization, where a coupling parameter $\beta < 0$ triggers instabilities in scalar perturbations, subsequent studies have shown that sufficiently large positive couplings can likewise destabilize the equilibrium configurations~\cite{Mendes:2014ufa,Mendes:2018qwo}. 
In static black holes undergoing spontaneous scalarization, however, such opposite-sign couplings typically inhibit the development of tachyonic instabilities.
By introducing angular momentum as a new degree of freedom into the theories, spin-induced spontaneous scalarization can extend the region of tachyonic instability into the regime of opposite-sign couplings, as demonstrated in ESGB gravity~\cite{Herdeiro:2020wei,Dima:2020yac} and EMS theory~\cite{Hod:2022txa,Lai:2022ppn}.
Further details can be found in review~\cite{Doneva:2022ewd}.  
On the other hand, beyond the framework of scalarization, the possibility of alternative mechanisms capable of generating hairy black holes remains not yet fully explored.

Motivated by these findings, we further investigate hairy black holes in EMS theory to identify hairy solutions that differ from those obtained via spontaneous scalarization.
To this end, we construct an EMS model featuring massless scalar field nonminimally coupled to the Maxwell field in asymptotically flat spacetime.
Adopting the exponential coupling function $f(\psi)$ as in~\cite{Herdeiro:2018wub}, we examine the existence of nontrivial scalar field behavior when the coupling parameter remains positive, subsequently analyzing the characteristic features of associated black hole solutions.
Building on our previous study of the EMS model with a massive real scalar field in asymptotically AdS spacetime where two types of hairy black hole states with distinct field profiles were identified~\cite{Guo:2024ymo}, the present study reveals a novel class of scalar hair controlled by the coupling parameter in flat spacetime.
The obtained scalar field exhibits monotonic radial growth while asymptotically approaching a constant value at spatial infinity.
In our earlier studies, tachyonic instability persists even for positive coupling values due to mass term contributions.
Here, however, the absence of scalar field mass in the current framework eliminates tachyonic instability for positive coupling, thereby establishing this solution as a fundamentally new class of hairy black hole distinct from spontaneous scalarization. 

The remainder of this paper is organized as follows.
In Sec.~\ref{sec=model}, we introduce the Einstein-Maxwell-Scalar model under consideration and derive the corresponding equations of motion along with the necessary boundary conditions.
Sec.~\ref{sec=solution} presents the numerical results for static black hole solutions, where we construct the hairy black hole configurations and explore their domain of existence as well as their key physical properties.
In Sec.~\ref{sec=perturb}, the stability analysis under scalar perturbation is conducted, which demonstrates the stability of the hairy black hole solutions.
Finally, Sec.~\ref{sec=conclusion} is devoted to further discussions and concluding remarks.

%%%%%%%%%%%%%%%%%%%%%%%%%%%Section%%%%%%%%%%%%%%%%%%%%%%%%%%%
\section{Einstein-Maxwell-Scalar Model}\label{sec=model}

In a 4-dimensional EMS theory, a massless real scalar field $\psi$ is nonminimally coupled to the Maxwell field via a coupling function $f(\psi)$, described by the action
\begin{equation}\label{eq=action}
	S=\int d x^4\left(R-\frac{1}{2} \nabla^\mu \psi \nabla_\mu \psi-\frac{1}{4} f(\psi) F_{\mu \nu}^2\right),
\end{equation}
where $R$ is the Ricci scalar, $\nabla_\mu$ denotes the covariant derivative, and $F_{\mu\nu}$ represents the electromagnetic field tensor.
The coupling function admits various possible parametrizations~\cite{Fernandes:2019rez}. 
In this work, we adopt an exponential quadratic coupling form~\cite{Herdeiro:2018wub}
\begin{equation}
	f(\psi)=e^{-\lambda \psi^2}
\end{equation}
with a coupling parameter $\lambda$.

From the action~\eqref{eq=action}, the scalar, electromagnetic, and gravitational fields in this model satisfy the following equations of motion
\begin{align}
   & \nabla_\mu \nabla^\mu \psi-\frac{\partial_\psi f}{4} F_{\mu \nu} F^{\mu \nu}=0, \label{eq=scalar}\\
   & \nabla^\mu\left(f(\psi)F_{\mu\nu}\right)=0,\label{eq=maxwell}\\
   & R_{\mu \nu}-\frac{1}{2} R g_{\mu \nu}=T_{\mu\nu}^{\psi}+f(\psi)T_{\mu\nu}^{\text{M}},\label{eq=metric}
\end{align}
where $\partial_\psi f=\frac{\partial f(\psi)}{\partial \psi}$.
The energy-momentum tensors for the scalar field and Maxwell field are given by
\begin{align}
	& T_{\mu\nu}^{\psi}=\frac{1}{2}\nabla_\mu \psi \nabla_\nu \psi-\frac{1}{4}g_{\mu\nu}\nabla_\rho\psi \nabla^\rho \psi,\\
	& T_{\mu\nu}^{\text{M}}=\frac{1}{2}F_{\mu \rho} F_\nu{ }^\rho-\frac{1}{8}g_{\mu\nu} F_{\sigma \rho}^2.
\end{align}
In what follows, we now proceed to construct hairy black hole solutions incorporating the backreaction of matter fields on spacetime geometry.
For this purpose, we employ the following {\it ansatz} for the corresponding fields
\begin{align}
	ds^2 & = - g(r)e^{-\chi(r)}dt^2 +\frac{dr^2}{g(r)} + r^2 (d\theta^2+\sin^2\theta d\varphi^2),\\
	A & = A_t(r)dt,\\
	\psi & = \psi(r).
\end{align}
The event horizon is located at $r_h$, defined by $g(r_h)=0$, while the asymptotic boundary corresponds to $r\rightarrow\infty$.
The hawking temperature of the black hole is expressed as~\cite{Cai:2022omk}
\begin{align}
	T_H=\frac{1}{4\pi}g'(r_h)e^{-\chi(r_h)/2}.
\end{align}

Substituting the {\it ansatz} into Eqs.~\eqref{eq=scalar}-\eqref{eq=metric}, we derive the following set of independent field equations
\begin{align}
	& \psi''(r) + \left(\frac{g'(r)}{g(r)}-\frac{\chi'(r)}{2}+\frac{2}{r}\right)\psi'(r) + \frac{e^{\chi(r)}A_t'(r)^2}{2 g(r)}f'(\psi)=0, \label{eq=motion1}\\
	& A_t''(r)+ \left(\frac{\chi '(r)}{2}+\frac{2}{r}+\frac{f'(\psi)}{f(\psi)}\psi'(r)\right)A_t'(r)=0, \label{eq=motion2}\\
	& \frac{g'(r)}{r} + g(r)\left(\frac{1}{r^2}+\frac{1}{4}\psi'(r)^2\right) + \frac{1}{4}f(\psi)e^{\chi(r)}A_t'(r)^2 - \frac{1}{r^2}=0, \label{eq=motion3}\\
	& 2 \chi '(r)+r \psi '(r)^2=0. \label{eq=motion4}
\end{align}
It's noteworthy that the electric charge $Q$ can be derived by the first integral of Eq.~\eqref{eq=motion2}
\begin{equation}\label{eq=charge}
	Q=-r^2f(\psi)e^{\chi/2}A_t'(r).
\end{equation}

To numerically solve the coupled system of ordinary differential equations for $\psi(r), A_t(r), g(r)$ and $\chi(r)$, it is convenient to introduce a compact coordinate defined by $z\equiv r_h/r$.
In this new radial coordinate, the above fields admit the following near-horizon expansions as $z\rightarrow 1$
\begin{align}
	\psi(z) &= \psi_h^0+\psi_h^1(1-z)+...,\label{eq=bdhorizon1}\\ 
	A_t(z)  &= a_h^1(1-z)+ a_h^2(1-z)^2+...,\label{eq=bdhorizon2}\\ 
	g(z)    &= g_h^1(1-z)+ g_h^2(1-z)^2+..., \label{eq=bdhorizon3}\\ 
	\chi(z) &= \chi_h^1(1-z)+.... \label{eq=bdhorizon4}
\end{align}
Substituting the above expansions into Eqs.~\eqref{eq=motion1}-\eqref{eq=motion4} and incorporating the constraint from Eq.~\eqref{eq=charge}, we determine the leading-order expansion coefficients as
\begin{align}
	a_h^1 &=-\frac{Q}{f(\psi_h^0)},
	& \psi_h^1 &=\frac{f'(\psi_h^0)}{f(\psi_h^0)}\frac{2Q^2}{Q^2-4f(\psi_h^0)}, \\
	g_h^1 &=1-\frac{Q^2}{4f(\psi_h^0)},
	& \chi_h^1 &=-\frac{f'(\psi_h^0)^2}{f(\psi_h^0)^2}\frac{2Q^4}{(Q^2-4f(\psi_h^0))^2}, 
\end{align}
expressed in terms of the fundamental parameters $\lambda, Q$ and $\psi_h^0$.

For the asymptotic boundary behavior, the field configurations in the far-field region take the forms
\begin{align}
    \psi(z\rightarrow 0) &=\psi_0 + \psi_1 z + ...,
    & A_t(z\rightarrow 0)&= \Phi - Q z +...,\\ 
	\chi(z\rightarrow 0) &= \chi_0 + \frac{\psi_1^2}{4}z^2 +...,  
	& g(z\rightarrow 0)  &= 1-2 M z +..., 
\end{align}
where $M$ corresponds to the ADM mass of the black hole, $\Phi$ denotes the chemical potential at spatial infinity, $\psi_0$ and $\psi_1$ parametrize the integration constants of the scalar field.

Before proceeding, we note that the equations of motion possess two distinct scaling symmetries
\begin{align}
    & t \to \lambda_t t, \quad e^{\chi} \to \lambda_t^2 e^{\chi}, \quad A_t \to \lambda_t^{-1} A_t; \\
    & r \to \lambda_r r, \quad g \to \lambda_r^2 g, \quad A_t \to \lambda_r A_t,
\end{align}
where $\lambda_t$ and $\lambda_r$ are dimensionless constants. 
The first symmetry allows us to fix $\chi(z\rightarrow 1)=\chi_h^0=0$ in the near-horizon boundary conditions without loss of generality.
The second symmetry permits us to normalize the horizon radius to $r_h=1$ for numerical convenience throughout subsequent calculations.

%%%%%%%%%%%%%%%%%%%%%%%%%%%Section%%%%%%%%%%%%%%%%%%%%%%%%%%%
\section{Numerical Solutions}\label{sec=solution}

Before proceeding to numerical computations, let us discuss how the sign of the coupling parameter influences the EMS model. 
According to the Klein–Gordon equation for the scalar field, the effective mass can be expressed as
\begin{equation}\label{eq=mass}
	m_{\mathrm{eff}}^2=\frac{1}{4} F_{\mu \nu}^2 \partial_\psi^2f(0)=\lambda\ e^{\chi(r)} A_t'(r)^2,
\end{equation}
which implies that the sign of the effective mass is entirely determined by the sign of the coupling parameter~\cite{Herdeiro:2018wub,Fernandes:2019rez}. 
Tachyonic instability requires a negative coupling, a scenario that has been thoroughly analyzed in the spontaneous scalarization of charged black holes.
Although a positive coupling indicates the absence of tachyonic instability, it does not preclude the existence of hairy black hole solutions beyond the RN configurations.

The system of field equations~\eqref{eq=motion1}-\eqref{eq=motion4} can be numerically integrated from the event horizon to the asymptotic region using the boundary conditions~\eqref{eq=bdhorizon1}-\eqref{eq=bdhorizon4}.
By employing the shooting method, we obtain numerical solutions for the field functions in the positive coupling regime ($\lambda>0$).
It is important to note that the leading asymptotic coefficient of the scalar field, $\psi_0$, is retained at spatial infinity.
When the coupling parameter $\lambda$ remains negative, the sign of the horizon expansion coefficient $\psi_h^1$ enforces monotonic decay of the scalar field from the event horizon to the asymptotic region~\cite{Herdeiro:2018wub,Fernandes:2019rez}. 
However, a positive coupling parameter reverses the sign of $\psi_h^1$, resulting in a monotonically increasing scalar field profile. 
Consequently, in this scenario, no solution exists with $\psi_0 = 0$. 
Nevertheless, it should be noted that asymptotically nonvanishing scalar field values are still allowed, even in scenarios involving spontaneous scalarization~\cite{Damour:1993hw,Belkhadria:2023ooc}.

%%%%%%%%%%%%%figure%%%%%%%%%%%%%%
\begin{figure}[thbp]
    \centering
    \includegraphics[height=0.31\linewidth]{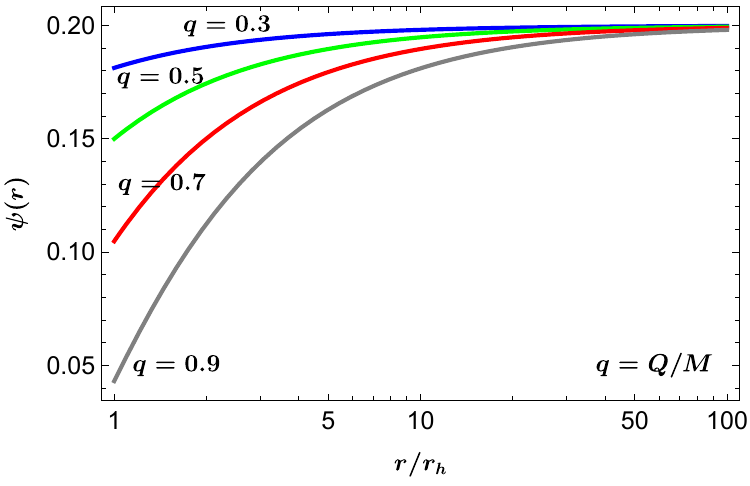}
    \includegraphics[height=0.31\linewidth]{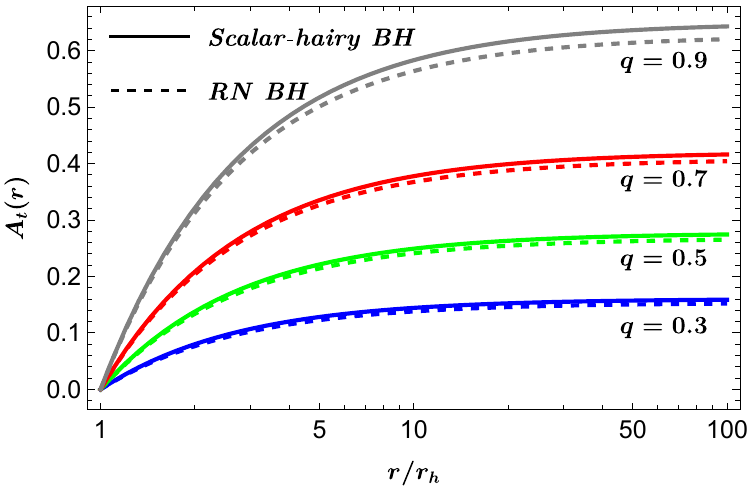}
    \includegraphics[height=0.31\linewidth]{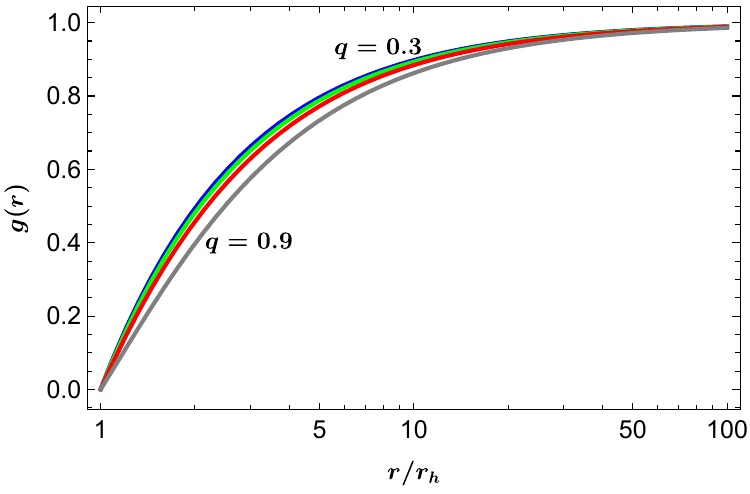}
    \includegraphics[height=0.31\linewidth]{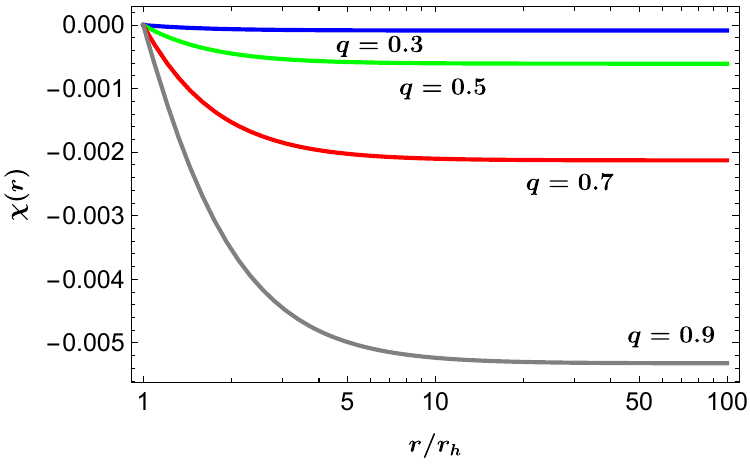}
    \caption{The profile of the field functions with $\lambda=1$ for different $q$.}
    \label{fig=profile}
\end{figure}
%%%%%%%%%%%%%%%%%%%%%%%%%%%%%%%%%

Our numerical analysis reveals that hairy black hole solutions persist across an extensive region of the fundamental parameter space $(\lambda,Q,\psi_h^0)$.
For convenience in our subsequent analysis, we employ the shooting method to obtain solutions for $\psi_0=0.2$ as an example, which allows us to investigate the profiles of the hairy solutions within a representative narrow parameter range. 

Let us begin by examining the profiles of hairy black hole solutions.
Fig.~\ref{fig=profile} illustrates the radial functions of various fields for coupling constant $\lambda = 1$ under different reduced charge $q = Q/M$. 
As mentioned earlier, the behavior of the scalar field is markedly different from that observed in spontaneous scalarization. 
It increases monotonically as one moves away from the event horizon and asymptotically approaches the constant $\psi_0$ at spatial infinity.
Notably, the scalar field profile steepens progressively with increasing $q$. 
Such a form of hairy black hole solution has also been reported in previous studies~\cite{Guo:2024ymo}, and hereafter we refer to these novel hairy black hole solutions as scalar-hairy black holes.

This observation is noteworthy because the behavior of the scalar field implies that the system does not converge to the RN black hole at asymptotic infinity. 
Analysis of the metric function $\chi(r)$ and the Maxwell field $A_t(r)$ reveals that enhanced deviations from the RN solutions accompany increasing $q$.
Remarkably, as $q$ approaches zero, the system is seen to revert to the uncharged Schwarzschild solution.
In the Schwarzschild case, the trivial scalar field solution possesses shift symmetry $\psi \rightarrow \psi + \mathcal{C}$ for any constant $\mathcal{C}$. 
When the Schwarzschild black hole is electrically charged, our scalar-hairy black holes can be interpreted as deformations of these shift-symmetric solutions.
Generally, the RN black hole can also be viewed as the charged counterpart of the Schwarzschild solution; however, in this case, the scalar field clearly remains trivial.
Spontaneous scalarization occurs due to the instability of the RN black hole, which leads to a very different scenario from our scalar-hairy solutions. 
These observations collectively indicate that the obtained scalar-hairy solutions constitute a distinct class from both scalarized and RN black holes in the EMS framework.

%%%%%%%%%%%%%figure%%%%%%%%%%%%%%
\begin{figure}[thbp]
    \centering
    \includegraphics[width=0.6\linewidth]{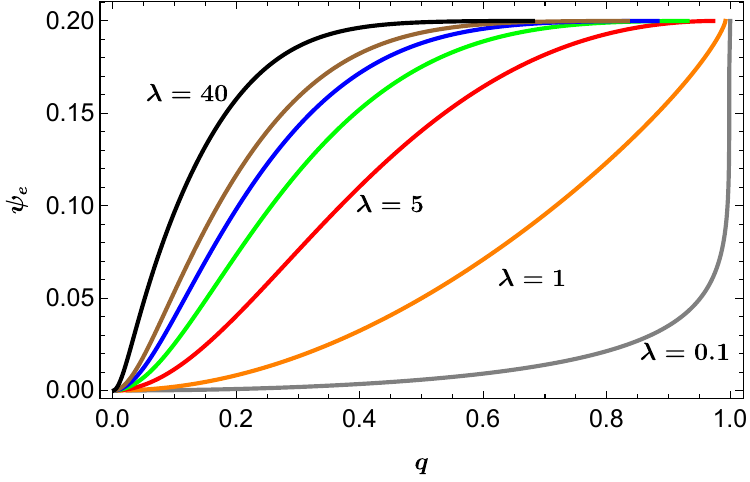}
    \caption{The dependence of the scalar charge $\psi_e\equiv\psi_0-\psi_h^0$ on the reduced charge $q$ for various coupling values. The curves, from lowest to highest, correspond to $\lambda=0.1,1,5,10,15,20,40$, respectively.}
    \label{fig=hair}
\end{figure}
%%%%%%%%%%%%%%%%%%%%%%%%%%%%%%%%%

Furthermore, the scalar and Maxwell fields exhibit remarkable similar characteristics in both their radial profiles and asymptotic boundary behaviors.
We denote the value of the scalar field at the event horizon $\psi_h^0$ by the scalar hair of the black hole, and define its deviation from the asymptotic value at infinity as $\psi_e\equiv\psi_0-\psi_h^0$, which quantifies the deviation of the field configuration from the trivial solution.
In our specific case with $\psi_0=0.2$, the trivial profile corresponds to $\psi_e=0$, while maximally nontrivial case achieves $\psi_e=0.2$.
By direct analogy with the Maxwell field, we therefore interpret $\psi_e$ as the scalar charge of the system, which measures the degree of nontriviality in the spatial profile of the scalar field configuration.
Fig.~\ref{fig=hair} presents the dependence of the scalar charge $\psi_e$ on the reduced charge $q$ for several values of the coupling $\lambda$.
In the limit $q\rightarrow 0$, $\psi_{e}$ vanishes for all $\lambda$, confirming the scalar field's trivial configuration in the neutral limit, with full gravitational recovery to the Schwarzschild solution when $Q=0$.
As $q$ increases, $\psi_{e}$ grows monotonically, saturating at its maximal value $\psi_e = 0.2$ before critical charge $q=1$.
Moreover, larger $\lambda$ yields a more rapid growth in $\psi_{e}$, which undergoes a sudden acceleration in the high-$q$ regime for weakly coupling case.
This trend indicates that increasing charge $q$ or coupling $\lambda$ favors the formation of scalar-hairy solutions that deviate further from the hairless state.

%%%%%%%%%%%%%figure%%%%%%%%%%%%%%
\begin{figure}[thbp]
    \centering
    \includegraphics[height=0.31\linewidth]{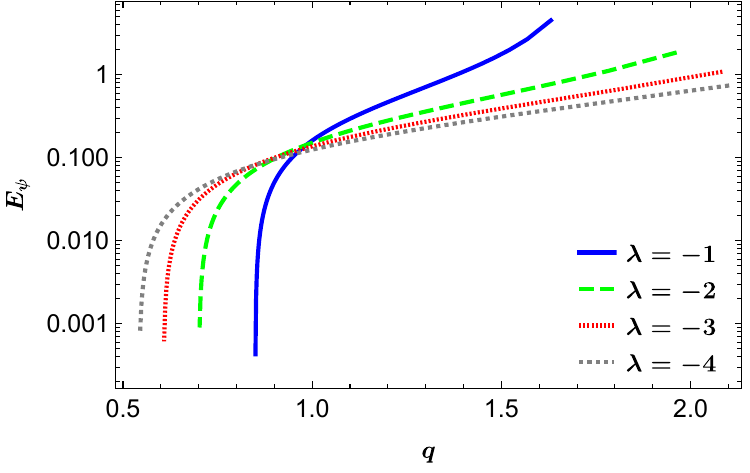}
    \includegraphics[height=0.31\linewidth]{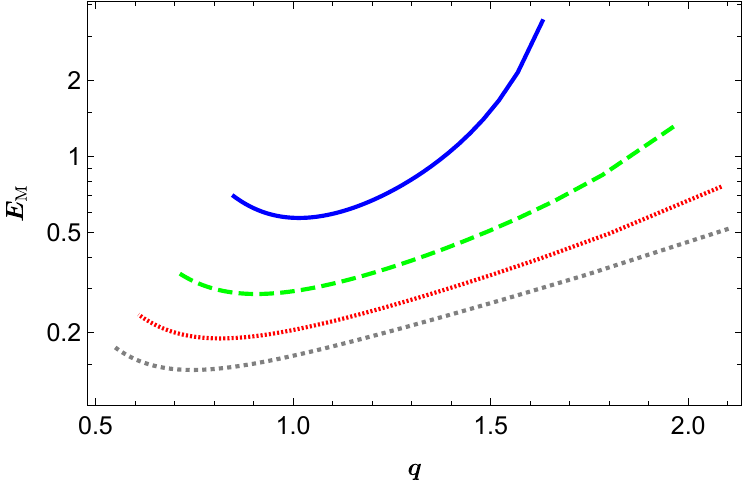}
    \includegraphics[height=0.31\linewidth]{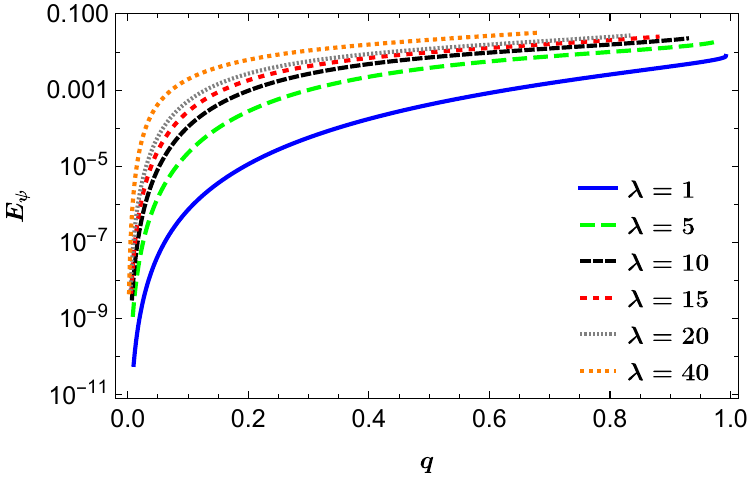}
    \includegraphics[height=0.31\linewidth]{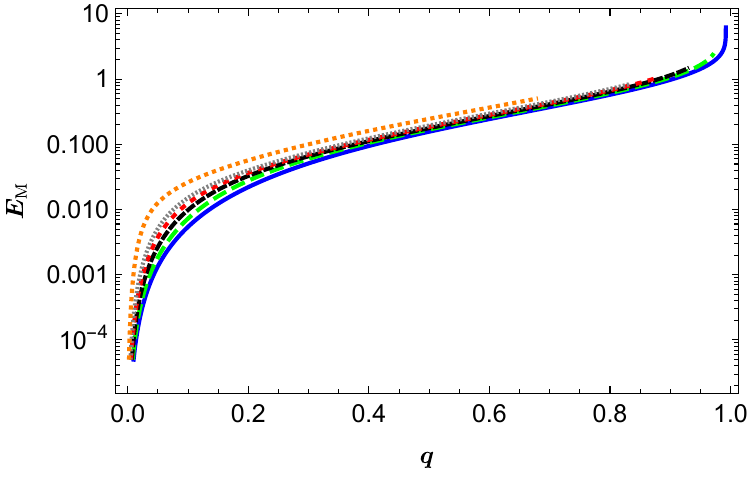}
    \caption{The energy of the scalar field and Maxwell field outside the horizon as the function of the charge $q$ for scalarization (top panel) and scalar-hairy solutions (bottom panel).}
    \label{fig=energy}
\end{figure}
%%%%%%%%%%%%%%%%%%%%%%%%%%%%%%%%%

To clarify the formation of scalar‑hairy black holes, we study the energy of the static solutions for scalar and electromagnetic fields, denoted by $E_{\psi}=\frac{1}{4\pi}\int_{r_h}^{\infty}T_{\mu\nu}^{\psi}n^{\mu}n^{\nu}dV$ and $E_{\text{M}}=\frac{1}{4\pi}\int_{r_h}^{\infty}T_{\mu\nu}^{\text{M}}n^{\mu}n^{\nu}dV$ on a normal vector of the constant time slice~\cite{Gourgoulhon:2007ue}.
Fig.~\ref{fig=energy} displays the relations between the field energy and the reduce charge $q$ for different coupling, where the upper panel corresponds to spontaneous scalarization ($\lambda<0$) and the lower panel represents the scalar-hairy solutions ($\lambda>0$).
During the charging process, both the scalar and maxwell fields accumulate energy, with the scalar field consistently storing less energy than its electromagnetic counterpart.
However, in the spontaneous scalarization case, one sees a nonmonotonic behavior in $E_{\text{M}}$: in the small-$q$ region, the electromagnetic field energy initially decreases before beginning to increase, indicating a temporary loss of small amounts of energy in the electromagnetic field at the onset of spontaneous scalarization.
Notably, increasing coupling strength in this case induces energy reduction in both scalar and Maxwell fields, whereas scalar-hairy black holes exhibit the opposite trend with enhanced energy storage at stronger coupling.
This intriguing observation of opposite energy transfer directions suggests that spontaneous scalarization and scalar-hairy black holes are driven by distinct mechanisms.
The spontaneous scalarization onset likely originates from tachyonic instability, while in positively coupled scalar-hairy black holes where tachyonic instability is absent, nonlinear coupling effects potentially dominate the energy dynamics.

\begin{figure}[thbp]
    \centering
    \includegraphics[width=0.6\linewidth]{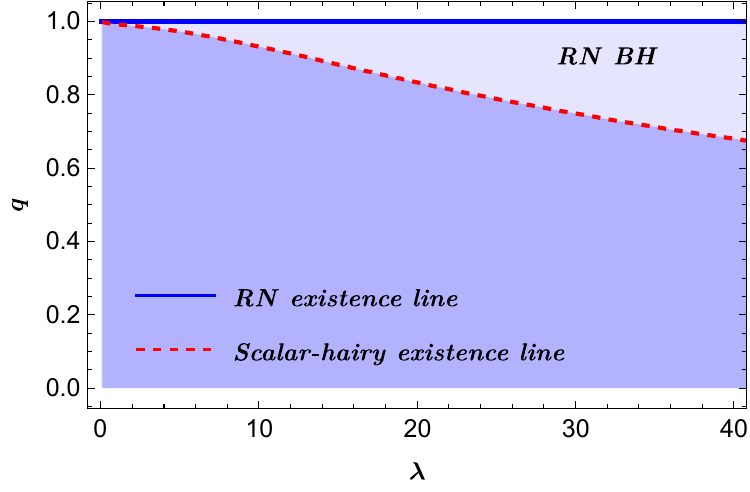}
    \caption{Existence domain of the scalar-hairy black holes in the parameter space of $(\lambda,q)$. }
    \label{fig=phase}
\end{figure}

Based on our numerical solutions obtained above, it is straightforward to delineate the existence domains of the scalar-hairy black holes in the parameter space $(\lambda,q)$.
As shown in Fig.~\ref{fig=phase}, the solid blue curve demarcates the RN black hole existence region, while the dashed red curve encloses the scalar-hairy solution parameter space.
Clearly, in contrast to the spontaneously scalarized black holes that persist in the $q>1$ domain~\cite{Herdeiro:2018wub,Fernandes:2019rez}, the scalar-hairy solutions are confined to $q<1$, entirely overlapping with the RN state.
This indicates that the scalar-hairy black holes are not the evolutionary endpoint of RN black holes but rather constitute competing gravitational phases.
When $q$ increases beyond the red dashed line, the electromagnetic effect becomes predominant, thereby driving the system toward RN black hole solutions.
The increase in the coupling parameter $\lambda$ reduces the viable range of $q$ for scalar-hairy black holes, a behavior consistent with the results presented in Fig.~\ref{fig=hair}.
In conjunction with Fig.~\ref{fig=phase}, this suggests that the existence line for scalar-hairy black holes may represent the onset point for RN black holes in EMS theory.

%%%%%%%%%%%%%figure%%%%%%%%%%%%%%
%%%%%%%%%%%%%%%%%%%%%%%%%%%%%%%%%

\begin{figure}[thbp]
    \centering
    \includegraphics[height=0.31\linewidth]{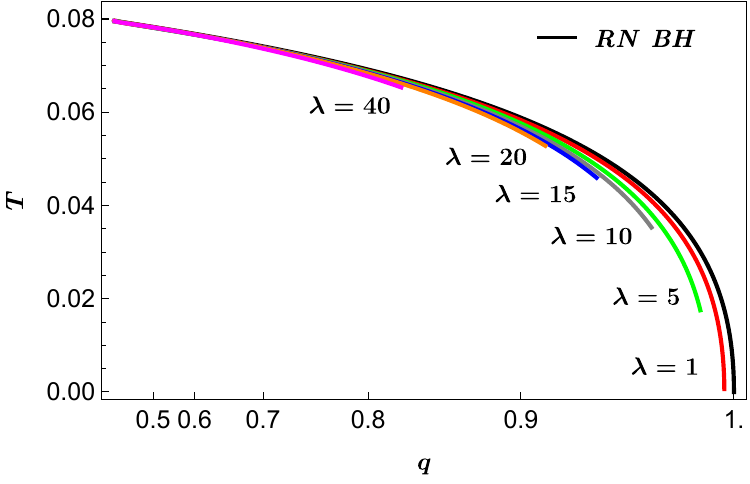}
    \includegraphics[height=0.31\linewidth]{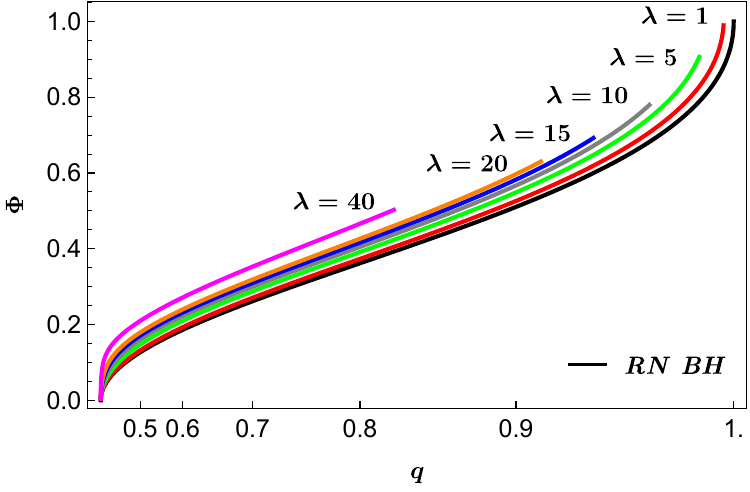}
    \caption{Left: Hawking temperature of the scalar-hairy black hole as a function of the reduced charge $q$ for different coupling value $\lambda$; Right: Chemical potential of the scalar-hairy solutions as a function of $q$ for different coupling value $\lambda$.}
    \label{fig=temperature}
\end{figure}

The behavior of the Hawking temperature as a function of $q$ for different coupling values is illustrated in the left panel of Fig.~\ref{fig=temperature}, with the $T-q$ relation for RN black holes represented by the black curve.
It is evident that the Hawking temperature of scalar-hairy black holes is consistently lower than that of RN black holes throughout their existence domain. 
For larger coupling values, temperature cannot drop to zero as $q$ increases.
In Fig.~\ref{fig=hair}, it was observed that scalar-hairy black holes transition smoothly into RN black holes with increasing charge.
However, the temperature profiles here suggest that this state transition is accompanied by a gap in temperature, although this gap diminishes with increasing coupling and eventually becomes negligible.

%%%%%%%%%%%%%figure%%%%%%%%%%%%%%
%%%%%%%%%%%%%%%%%%%%%%%%%%%%%%%%%

Notably, due to the nonminimal coupling between the electromagnetic and scalar fields, scalar-hairy black holes also carry electric charge. 
The right panel of Fig.~\ref{fig=temperature} reveals the chemical potential $\Phi$ as the function of $q$.
Compared to the RN black hole, the scalar-hairy black hole tends to develop a larger chemical potential, which is consistent with the profile behavior of the Maxwell field $A_t(r)$ in Fig.~\ref{fig=profile}.
Although an increase in the coupling parameter leads to a reduction in the effective interaction strength, the figure shows that the resulting scalar-hairy black hole solutions deviate further from the RN state. 
Additionally, higher coupling values constrain the charge domain of the black hole, making it more difficult for the scalar-hairy solutions to acquire charge.
Consequently, when the black hole becomes overcharged, the scalar-hairy solutions may become unstable, triggering a transition to the RN black hole state, accompanied by discontinuities in both the Hawking temperature and electromagnetic chemical potential.

%%%%%%%%%%%%%%%%%%%%%%%%%%%Section%%%%%%%%%%%%%%%%%%%%%%%%%%%
\section{Perturbation Stability}\label{sec=perturb}

In order to study the stability of the scalar-hairy black hole solutions against radial perturbation, we consider the linear perturbation of the field functions
\begin{align}
	& ds^2 = (g(r)+\epsilon g_p(t,r))e^{-\chi(r)+\epsilon \chi_p(t,r)}dt^2 +\frac{dr^2}{g(r)+\epsilon g_p(t,r)} + r^2 (d\theta^2+\sin^2\theta d\varphi^2),\\
	& \psi(t,r)=\psi(r)+\epsilon \frac{\Psi(t,r)}{r},\quad A_t(t,r)=A_t(r)+\epsilon \delta A_t(t,r),
\end{align}
where $g(r),\ \chi(r),\ \psi(r)$ and $A_{t}(r)$ are the background numerical solutions of the scalar-hairy black hole obtained in Sec.~\ref{sec=solution}, while $g_p(t,r),\ \chi_p(t,r),\ \Psi(t,r)$ and $\delta A_t(t,r)$ represent the perturbed fields around the scalar-hairy black hole background. $\epsilon$ denotes the control parameter of perturbations. 

After substituting the perturbations into the field equations~\eqref{eq=scalar}-\eqref{eq=metric}, at first order in $\epsilon$, we have the relations
\begin{align}
	& g_p(r)=-\frac{1}{2}g(r)\psi'(r)\Psi(r),\\
	& \chi_p(r)=\left(\frac{\Psi(r)}{r}-\Psi'(r)\right)\psi'(r),\\
	& \delta A_t'(r)=-\left(\frac{1}{2}f(\psi)\chi_p(r)+\frac{\Psi(r)}{r}\partial_{\psi}f(\psi)\right)\frac{A_t'(r)}{f(\psi)}.
\end{align}
This indicates that all of the metric and electromagnetic perturbations decouple completely from the linearized system, leaving scalar field perturbation as the sole dynamical degree of freedom.
In the following discussion, we focus on the analysis of spherically symmetric modes ($\ell=0$). 
Admitting a further decomposition $\Psi(t,r)=\phi(r)e^{-i\omega t}$, the linearized scalar perturbation satisfies a Schr\"odinger-type equation
\begin{equation}\label{eq=schr}
    \frac{d^2 \phi(r)}{d r_{*}^2}+\left(\omega^2-V_\mathrm{eff}(r)\right) \phi(r)=0 ,
\end{equation}
where the tortoise coordinate is defined by $r_*=\int \frac{d r}{g(r)e^{-\chi(r)/2}}$.
The effective potential takes the form
\begin{equation}\label{eq=poten}
    V_\mathrm{eff}(r)=\frac{e^{-\chi(r)}}{r^2}g(r)\left(1-g(r)-\frac{1}{2}r^2\psi'(r)^2-r^2A_t'(r)^2e^{-\lambda\psi(r)^2}e^{\chi(r)}G(r)\right),
\end{equation}
with
\begin{equation}\label{eq=potenG}
	G(r)=\frac{1}{4}-\lambda-r\lambda\psi(r)\psi'(r)-\frac{1}{8}r^2\psi'(r)^2-2\lambda^2\psi(r)^2.
\end{equation}

%%%%%%%%%%%%%figure%%%%%%%%%%%%%%
\begin{figure}[thbp]
    \centering
    \includegraphics[height=0.31\linewidth]{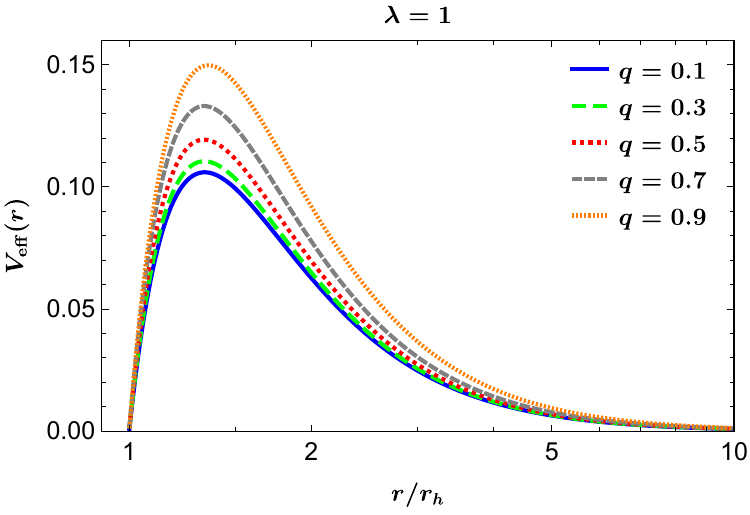}
    \includegraphics[height=0.31\linewidth]{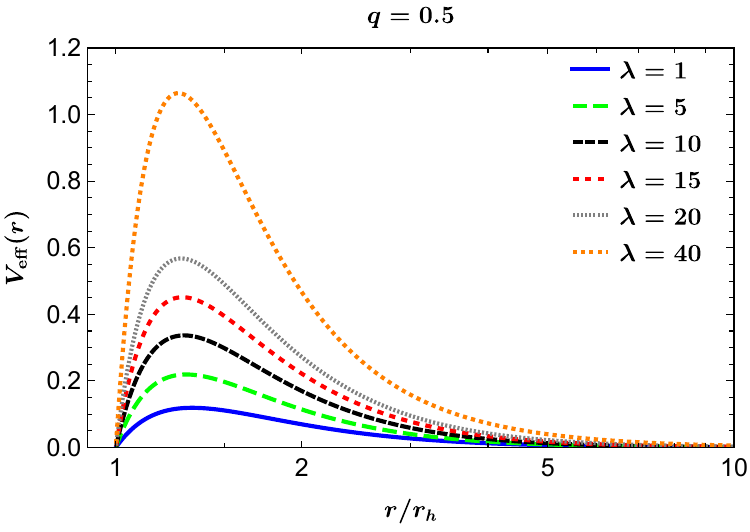}
    \caption{Left: the profiles of the effective potential for different reduced charge $q$ with $\lambda=1$; Right: the profile of the effective potential for different value of the coupling parameter $\lambda$ with $q=0.5$.}
    \label{fig=potenq}
\end{figure}
%%%%%%%%%%%%%%%%%%%%%%%%%%%%%%%%%

Herein, we analyze the behavior of effective potentials as shown in Fig.~\ref{fig=potenq}. 
The scalar perturbation forms a positive potential barrier near the event horizon, with the barrier height increasing both as $q$ or $\lambda$ rises.
These behaviors are markedly different from that observed in spontaneous scalarization with negative coupling~\cite{Fernandes:2019rez}, where the potential develops negative wells near the RN limit that could trigger instabilities. 
The positivity of the effective potential indicates that Eq.~\eqref{eq=schr} will not have a bound state, which will lead to unstable modes with $\omega^2<0$. 
Therefore, this behavior ensures the stability of our scalar-hairy black hole solutions against scalar perturbation.

The stability can be further examined by calculating the QNMs of the perturbation equation.
Following our analysis of the effective potential, scalar perturbation satisfies boundary conditions corresponding to purely ingoing waves near the horizon and purely outgoing waves at spatial infinity.
The computed $\omega$ represents the complex discrete frequencies, where the real part corresponds to the oscillation frequency while the imaginary part characterizes the mode's (un)damping rate.
A positive imaginary component indicates an exponentially growing perturbation, signaling an instability in the black hole background.
Using the numerical background solutions obtained in the previous section, we compute the fundamental QNM frequencies by employing the asymptotic iteration method (AIM)~\cite{Cho:2011sf}.

%%%%%%%%%%%%%figure%%%%%%%%%%%%%%
\begin{figure}[thbp]
    \centering
    \includegraphics[height=0.31\linewidth]{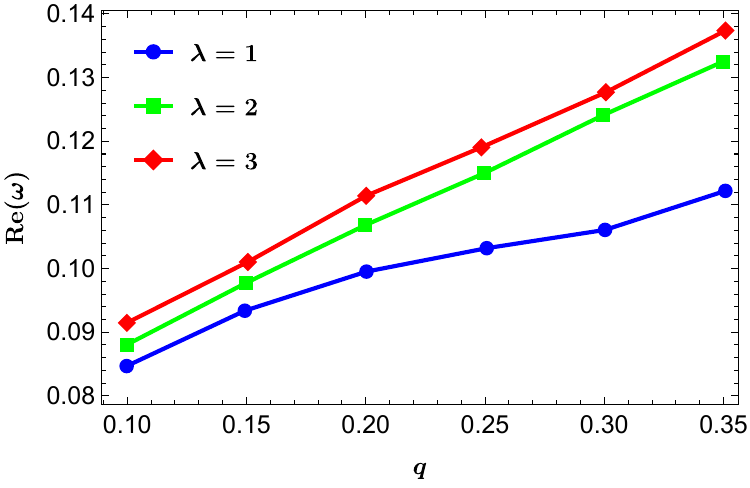}
    \includegraphics[height=0.31\linewidth]{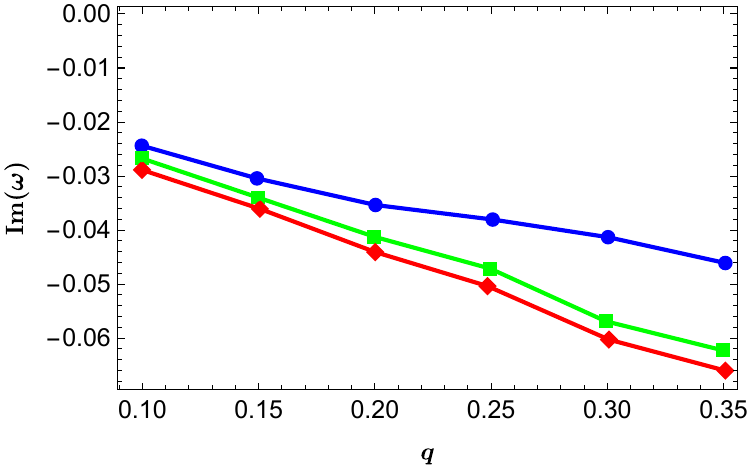}
    \caption{The fundamental QNM frequencies of the scalar perturbation as a function of $q$ with different coupling.}
    \label{fig=qnm}
\end{figure}
%%%%%%%%%%%%%%%%%%%%%%%%%%%%%%%%%

The computational challenges in analyzing scalar perturbations arise from the numerical background solutions and associated convergence difficulties inherent to the system.
As illustrative examples, we calculate several fundamental QNM frequencies at small coupling $\lambda$ and charge $q$ in Fig.~\ref{fig=qnm}.
Both the magnitude of real and imaginary parts of QNMs exhibit monotonic growth with increasing $q$ and $\lambda$.
Specifically, the enhancement of $\text{Re}(\omega)$ indicates progressively stronger oscillatory behavior in perturbations, while the increasing magnitude of $\text{Im}(\omega)$ corresponds to a faster decay rate.
These results further demonstrate linear stability for these scalar-hairy black hole solutions under scalar perturbation.

%%%%%%%%%%%%%%%%%%%%%%%%%%%Section%%%%%%%%%%%%%%%%%%%%%%%%%%%
\section{Results and further discussion}\label{sec=conclusion}

In this work, we examine black hole solutions with scalar hair in the EMS theory in the 4-dimensional asymptotically flat spacetime.
It is well-established that EMS theory exhibits spontaneous scalarization of black hole~\cite{Herdeiro:2018wub,Fernandes:2019rez}, where a scalar field with a negative coupling parameter induces tachyonic instabilities in the original RN spacetime.
However, our analysis reveals when the coupling parameter keeps positive, the nonlinear coupling can still lead to hairy black hole solutions that are distinct from those produced by spontaneous scalarization.

In contrast to the negative coupling case where the scalar field decays with increasing radial coordinate, a defining characteristic of these scalar-hairy black holes is the monotonic growth of the scalar field along the radial direction, asymptotically approaching a constant at infinity.
Similar behavior for the scalar field profile has also been observed in the EMS theory formulated in an asymptotically AdS background~\cite{Guo:2024ymo}.
This phenomenon arises from the boundary conditions imposed on the scalar field near the event horizon, where the coupling parameter's sign determines the sign of the first-derivative term in Eq.~\eqref{eq=bdhorizon1}, governing scalar field evolution near the horizon.
Furthermore, from another perspective, the scalar-hairy black hole solutions can be regarded as the charged counterparts of the Schwarzschild black hole.
Although the RN black hole also represents a charged Schwarzschild solution, it does not support scalar hair in the EMS theory, which distinguishes it from our scalar-hairy black holes.

We also observe a striking contrast between the spontaneous scalarization and scalar-hairy black hole branches in the energies of the static field solutions.
In spontaneous scalarization, both scalar and electromagnetic field energies are suppressed by increasing the coupling value, whereas in scalar-hairy black holes they are enhanced with stronger coupling.
This reversal of energy transfer clearly shows that scalar-hairy black holes do not originate from the conventional mechanism of spontaneous scalarization.
The onset of spontaneous scalarization requires sufficient coupling magnitude or charge to generate tachyonic instabilities through sufficiently negative effective mass term.
However, there is no evidence that increasing coupling magnitude beyond this onset drives the solution further away from the trivial profile.
In fact, some studies have shown that scalar hair decreases with increasing coupling amplitude after reaching a saturation value~\cite{Myung:2018jvi,Guo:2020sdu}.
By contrast, in scalar-hairy black holes, nonlinear coupling plays the dominant role, meaning that the coupling amplitude is positively correlated with the degree of deviation from the trivial solution.

Numerical analysis reveals competitive interplay between scalar and electromagnetic fields for the scalar-hairy black holes.
An increase in the charge $q$ causes the scalar-hairy black hole solutions to deviate further from the RN black hole, though overcharging drives the system back to the electromagnetic vacuum state.
Contrary to spontaneous scalarization, where enhanced coupling enlarges the existence domain of scalarized black holes, our results show that an increase in $\lambda$ compresses the region admitting the scalar-hairy black hole state.
Furthermore, a positive coupling parameter effectively weakens the interaction strength.
As $\lambda\rightarrow\infty$, the system reduces to Einstein-Hilbert gravity with a massless scalar, where no stable hairy solutions exist~\cite{Israel:1967wq}. 
Consequently, as shown in Fig.~\ref{fig=phase}, the scalar-hairy black hole state occupies a parameter subspace within the RN boundary, whereas in the scalarization case an increased coupling can drive the scalarized branch into the $q>1$ regime. 
Combined with the results of Fig.~\ref{fig=hair}, this indicates that the scalar-hairy black holes cannot be viewed as the endpoint of the RN black holes but might serve as a potential starting point for RN evolution in the EMS theory. 
Finally, our stability analysis reveals that the effective potential for scalar perturbations on the numerical background is positive, and that the associated QNMs we obtained have negative imaginary parts.
These results confirm that our scalar-hairy solutions as a new branch of hairy black holes distinct from RN family, remain linearly stable at least against scalar perturbations.
A comprehensive stability assessment of these scalar‑hairy black holes requires further examination under either gravitational or electromagnetic perturbations~\cite{Zou:2020zxq,Blazquez-Salcedo:2022omw, Blazquez-Salcedo:2018jnn}.
Furthermore, a more detailed discussion of the thermodynamic examination and interrelations of scalar-hairy and RN black holes in the vicinity of the existence line will further elucidate their properties and the potential black hole phase transitions.

%%%%%%%%%%%%%%%%%%%%%%%%%%%%%%%%%%%%%%%%%%%%%%%%%%%%%%%%%%%%%%%%%%%%%%%%%%%%
\begin{acknowledgments}
We thank Dr. Hang Liu and Ruan-Ru Li for their helpful discussion.
This work is supported in part by the financial support from Brazilian agencies 
Funda\c{c}\~ao de Amparo \`a Pesquisa do Estado de S\~ao Paulo (FAPESP), 
Fundação de Amparo à Pesquisa do Estado do Rio Grande do Sul (FAPERGS),
Funda\c{c}\~ao de Amparo \`a Pesquisa do Estado do Rio de Janeiro (FAPERJ), 
Conselho Nacional de Desenvolvimento Cient\'{\i}fico e Tecnol\'ogico (CNPq), 
and Coordena\c{c}\~ao de Aperfei\c{c}oamento de Pessoal de N\'ivel Superior (CAPES).

\end{acknowledgments}

%%%%%%%%%%%%%%%%%%%%%%%%%%%%%%%
\bibliographystyle{jhep}
\bibliography{refs}
\end{document}